# Two-slit Experiment And Wave-Particle Duality

N. L. Chuprikov

*Department of Theoretical Physics, Tomsk State Pedagogical University, Tomsk, Russia*

**Abstract.** On the basis of an alternative approach to micro-cat states (Found. of Phys., 41, No. 9, p.1502 (2011)) we develop a new model of the two-slit experiment. It explains both this particular experiment and how the wave properties of any ensemble of single quantum particles emerge from their corpuscular ones. The key role in this explanation is played by the indistinguishability of identical quantum particles.

**Keywords:** Wave-particle duality. Statistical interpretation. Indistinguishability of identical quantum particles. Bohmian trajectories.

## INTRODUCTION

The main peculiarity of the well-known two-slit experiment is that it deals with a one-particle quantum phenomenon which must be treated, from the viewpoint of classical probability theory, as a complex process consisting of two alternative subprocesses. For a point-like classical particle can pass only through one of two open slits. However, by quantum mechanics the state of a particle in this process represents the micro-cat state whose substates interfere with each other and cannot be associated with alternative subprocesses of the two-slit experiment − within the contemporary approach to micro-cat states the superposition principle contradicts the "either-or" rule which governs mutually exclusive random events in classical probability theory.

By Bohr's complementarity principle the wave and corpuscular properties of a quantum particle are mutually incompatible − there is no experimental setup which would allow their simultaneous measurement. Later (see [1-3] and references therein) this principle has been, in fact, replaced by a mathematical theorem to have corrected Bohr's statement about the wave-particle duality. Unlike Bohr's principle the deduced in [1-3] mathematical equation, which connects the visibility of the interference pattern with the distinguishability of the path of propagation of a particle, allows "unsharp" (but accurate) joint measurements of the wave and corpuscular properties of a particle. However, it forbids their "sharp" joint measurements. Thus, in fact these studies maintain Bohr's principle and confirm once more that the quantum-mechanical superposition principle conflicts with classical "either-or" rule.

Nevertheless, the studies [1-3] are important for resolving this conflict, as they change its status. Indeed, they evidence that the incompatibility of the wave and corpuscular properties of a quantum particle in the two-slit experiment is merely a consequence of the contemporary quantum-mechanical treatment of the micro-cat state to describe this experiment, rather than an unambiguously established physical principle. Thus, there is something wrong in this treatment, because it opposes the superposition principle to probability theory and endows a quantum particle with mutually incompatible physical properties.

In order to reveal this "something", we have to attract reader's attention to Bohr's statement concerning quantum phenomena: "*the unambiguous interpretation of any measurement must be essentially framed in terms of the classical physical theories*" [4]. This means that statistical experimental data, being obtained for any quantum phenomenon with the help of a *macroscopic* device, are under jurisdiction of *classical* probability theory. And those approaches to study quantum phenomena from the viewpoint of this theory are of great importance for revealing discrepancies to exist at present between quantum and classical probabilities to describe micro-cat states.

For example, as was shown in [5], the set of statistical data described by the squared modulus of the micro-cat state in the two-slit experiment belongs to a non-Kolmogorovian probability space, what means (see [6]) that such data are mutually incompatible. This concerns all quantum phenomena where micro-cat states appear, and namely the disregard to this peculiarity of micro-cat states leads to paradoxes (e.g., the Cat and Hartman paradoxes) in the contemporary quantum-mechanical models of such phenomena.

By probability theory the squared modulus of a micro-cat state cannot be treated as probability density and the unambiguous interpretation of experimental data associated with such state implies decomposing the probability space to underlie its squared modulus into the sum of Kolmogorovian ones. The alternative approach [7] to micro-cat states presents such decomposition for a 1D completed scattering and two-slit diffraction (see also [8]). Here we dwell once more on the two-slit experiment, highlighting and developing further the key points of its model.

## TWO-SLIT DIFFRACTION AS A COMPLEX PROCESS CONSISTING OF TWO ALTERNATIVE SUBPROCESSES

Following [7,8], let us consider (in two dimensions) the strictly symmetrical setting of the two-slit experiment, assuming that the $y$-axis coincides with the first screen to have two parallel identical slits centered on the lines $y = -a$ (first slit) and $y = a$ (second slit). A particle with a given energy $E$ impinges the first screen from the left; the second screen coincides with the line $y = L$ ($L > 0$). Let also $\Psi_{one}(x, y; E)$ be the "one-slit" wave function to describe the state of a particle in the one-slit experiment; $\Psi_{one}(x, -y; E) = \Psi_{one}(x, y; E)$; $\Psi_{one}(0, y; E) = 0$, if $|y| > 0$. Then the wave function $\Psi_{two}(x, y; E)$ to describe the two-slit experiment is

$$\Psi_{two}(x, y; E) = \Phi(x, y; E); \qquad (1)$$

$$\Phi(x, y; E) = \psi_{one}^{(1)}(x, y; E) + \psi_{one}^{(2)}(x, y; E);$$

$$\psi_{one}^{(1)}(x, y; E) = \Psi_{one}(x, y + a; E), \quad \psi_{one}^{(2)}(x, y; E) = \Psi_{one}(x, y - a; E).$$

The micro-cat state (1) is the superposition of the one-slit substates $\psi_{one}^{(1)}$ and $\psi_{one}^{(2)}$ connected to different macroscopic slits and, by classical physics, they must describe alternative subprocesses. However this is not the case, as $\|\Psi_{two}\|^2 \neq \|\psi_{one}^{(1)}\| + \|\psi_{one}^{(2)}\|$ because of interference between these substates; note that for any wave function $\varphi$ its norm reads as $\|\varphi\|^2 = \int_{-\infty}^{\infty} |\varphi(L, y; E)|^2 dy$. By probability theory the statistical data described by the $y$-distribution $|\Psi_{two}(L, y; E)|^2$ are mutually incompatible, because they belong to a non-Kolmogorovian probability space [5,6].

The fact that the macroscopically distinct substates $\psi_{one}^{(1)}$ and $\psi_{one}^{(2)}$ cannot be associated with alternative subprocesses of the two-slit diffraction is usually treated as the evidence of incompatibility of the wave and corpuscular properties of a quantum particle. However, by the approach [7], this is not the case, as the mentioned non-Kolmogorovian probability space is decomposable into the sum of Kolmogorovian ones to describe alternative subprocesses of the two-slit diffraction (see also [8]). As was shown in [7,8], the state $\Psi_{two}$ can be rewritten in the form

$$\Psi_{two}(x, y; E) = \psi_{two}^{(1)}(x, y; E) + \psi_{two}^{(2)}(x, y; E); \qquad (2)$$

$$\psi_{two}^{(1)}(x, y; E) = \begin{cases} 0, & y > 0 \\ \Phi(x, y; E), & y < 0 \end{cases}; \qquad \psi_{two}^{(2)}(x, y; E) = \begin{cases} \Phi(x, y; E), & y > 0 \\ 0, & y < 0 \end{cases}$$

$\psi_{two}^{(1)}(x, 0; E) = \psi_{two}^{(2)}(x, 0; E) = \Phi(x, 0; E)/2$. Here $\psi_{two}^{(1)}$ and $\psi_{two}^{(2)}$ describe alternative subprocesses, because $\|\Psi_{two}\|^2 = \|\psi_{two}^{(1)}\| + \|\psi_{two}^{(2)}\|$. The physical ideas to underlie the transformation of the micro-cat state (1) into (2) are as follows.

The $\mathcal{P}$-context (see [6,7]) to generate the ensemble of particles taking part in the two-slit experiment is a complex one to consist of two elementary $\mathcal{P}$-contexts; by [7] each slit generates its own elementary $\mathcal{P}$-context and corresponding subensemble with the fixed number of particles. Both subensembles evolve in the non-overlapping spatial regions separated by a boundary to coincide, in the considered setting of the two-slit experiment, with the symmetry line $y = 0$.

By quantum mechanics particles never cross this line, because the $y$-projection of the probability current density to correspond $\Psi_{two}$ is zero on this line. That is, in the two-slit experiment the symmetry line $y = 0$ "acts" on the *ensemble* of particles as a infinitesimally thin two-side mirror to elastically reflect particles. If one really inserted a mirror along this line, the interference pattern on the second screen remained the same for any distance $L$.

Thus, the ensemble of particles, freely moving between the first and second screens in the original two-slit experiment (without the mirror), is *equivalent* to the ensemble of particles in the modified experiment (with the mirror) where particles *a priori* cannot cross the line y = 0 occupied by the mirror. As a result, the original ensemble of particles taking part in the two-slit experiment falls into two subensembles: one of them, described by $\psi_{two}^{(1)}$, consists of particles to pass through the first slit in the screen (provided that the second one is open); another, described by $\psi_{two}^{(2)}$, consists of particles to pass through the second slit (provided that the first one is open). That is, by this approach a quantum particle like classical one can pass only through one of two open slits. In this sense, the above explanation partially reconciles the wave and corpuscular properties of a quantum particle.

However, their full reconciliation must also explain why the (immaterial) symmetry line y = 0, in the original two-slit experiment, "acts" on the *ensemble* of particles as an ideally impenetrable (material) mirror. Indeed, this wave property of the ensemble seems to be incompatible with the corpuscular properties of its single members.

## WAVE-PARTICLE DUALITY AND INDISTINGUISHABILITY OF PARTICLES IN QUANTUM ENSEMBLE

Our explanation will be based on the idea that a quantum particle, as a (probably not point-like, but) well-localized on the atomic scales object, moves along some random trajectories continuously evolved in the space-time. This idea is relevant since, in line with the statistical interpretation of quantum mechanics, this theory is incomplete and its statistical dispersion principle does not forbid the existence of one-particle trajectories. The continuity of these trajectories and the causality principle are sufficient to say that the above inference that "the ensemble of particles in the original two-slit experiment is *equivalent* to that in the modified experiment" does not at all mean that the trajectories of particles in the original ensemble do not cross the symmetry line. In the original two-slit experiment we deal in fact with the ensemble of paired dissymmetrical trajectories; one trajectory in each pair is described by $\psi_{one}^{(1)}$, but another to pass through the second slit is described by $\psi_{one}^{(2)}$.

Let us consider those paired trajectories which cross the symmetry line. In each such pair the trajectories intersect each other, and the point of their intersection lies just on thy symmetry line. Since the total "flux" through this line is zero in the pair, in virtue of the indistinguishability of particles at the intersection points, each pair of intersected trajectories is equivalent to (indistinguishable from) the pair of trajectories to be tangent to each other on the symmetry line − one trajectory of this pair, which passes through the first slit is described by $\psi_{two}^{(1)}$; another one to pass through the second slit is described by $\psi_{two}^{(2)}$.

Note that this explanation gives us something more than the sought answer on the above particular question. As is seen, the key role in this explanation is played by the quantum-mechanical principle of *indistinguishability* of identical particles. In the same experiment with *classical* particles, the pair of intersected dissymmetrical one-particle trajectories is nonequivalent to the pair of tangent ones because of the *distinguishability* of identical classical particles at the points of intersection. This gives grounds to assume that *the whole wave dynamics of quantum one-particle ensembles is inseparably linked with the indistinguishability of its single members.*

Indeed, one of the main points of the above particular explanation is that the number of particles in each of the two regions $y \leq 0$ and $y \geq 0$ is constant in time, because the "number" of trajectories coming to the symmetry line is equal to that of outgoing trajectories. However, this line is just a *current* line (to be common for $\psi_{two}^{(1)}$ and $\psi_{two}^{(2)}$). Thus, we can assume that any other current line in each region possesses the same properties − it can be considered as the set of intersection points for paired trajectories which create zero total flux through this line.

Thus again, due to indistinguishability of quantum particles at the points of intersection, the original set of paired one-particle trajectories to intersect this current line is equivalent to the set of paired trajectories, tangent at this line. Then, since this is valid for any probability current line and since the set of such lines is everywhere dense in the domain of a given probability wave, we arrive at the conclusion that the original set of mutually intersecting trajectories to describe the corpuscular properties of particles in the ensemble is equivalent to (indistinguishable from) the set of probability current lines to describe their wave dynamics.

As is seen, we have led to the conception of Bohmian trajectories to coincide with probability current lines. That is, in fact, our approach to micro-cat states, which implies the existence of local hidden variables, supports the nonlocal Bohmian approach − it reveals the physical sense of Bohmian trajectories. Namely, it says that, in the general case, a single Bohmian trajectory does not represent a real trajectory of a quantum particle in a single experiment; however the *ensemble* of Bohmian trajectories is always equivalent to (indistinguishable from) the corresponding ensemble of real one-particle trajectories.

Thus, the Bohmian approach is not a prequantum theory whose existence is implied by the statistical interpretation of quantum mechanics. Rather it is a tool of calculating the probability current lines to have the above physical meaning. Of importance is also to stress that, in the case of a micro-cat state, Bohmian trajectories can be defined only for its substates to describe alternative subprocesses. For example, in the case of a 1D completed scattering, the known Bohmian trajectories must be recalculated on the basis of the wave functions to describe the transmission and reflection subprocesses presented, e.g., in [7].

So, our approach not only explains the two-slit experiment, it also prompts the idea − there is something in the (hidden) structure of a quantum particle, which makes it indistinguishable from another identical particle and, simultaneously, accounts for its wave properties.

## ACKNOWLEDGMENTS

This work was supported (in part) by Russian Science and Innovations Federal Agency under contract No 02.740.11.0238 as well as by the Programm of supporting the leading scientific schools of RF (grant No 3558.2010.2).

## REFERENCES

1. W. K. Wootters and W. H. Zurek, P, *Phys. Rev. D,* **19**, 473-484 (1978).
2. P. Mittelstaedt, A. Prieur, and R. Schieder, *Found. of Phys.,* **17**, 891-903 (1987).
3. G. Jaeger, A. Shimony, and L. Vaidman, *Phys. Rev. A,* **51**, 54-67 (1995).
4. N. Bohr, *Nature,* **128**, 691-692 (1931).
5. L. Accardi, *Vestn. of Samar. State Univer.: Natur. Science Series*, No 8/1 (67), 277-294 (2008).
6. A. Yu. Khrennikov, *Found. of Phys.*, **35**, No. 10, 1655-1693 (2005).
7. N. L. Chuprikov, *Found. of Phys.*, **41**, No. 9, 1502-1520 (2011).
8. N. L. Chuprikov, *Vestn. of Samar. State Univer.: Natur. Science Series*, No 2 (23), 235-242 (2011).